\documentclass[conference]{IEEEtran}
\pdfoutput=1
\usepackage{pdfpages}
\usepackage{blindtext, graphicx}
\usepackage{todonotes}
\usepackage{hyperref}

\ifCLASSINFOpdf
\else
\fi
\IEEEoverridecommandlockouts

\begin{document}
%
\title{Unsupervised video summarization framework using keyframe extraction and video skimming}


\author{\IEEEauthorblockN{Shruti Jadon \thanks{This work has been done as part of 670 Computer Vision Coursework at UMass Amherst.}}
\IEEEauthorblockA{College of Information and Computer Science\\
University of Massachusetts\\
Amherst, MA 01002\\
Email: sjadon@umass.edu}
\and
\IEEEauthorblockN{Mahmood Jasim}
\IEEEauthorblockA{College of Information and Computer Science\\
University of Massachusetts\\
Amherst, MA 01002\\
Email: mjasim@cs.umass.edu}}
\maketitle

\begin{abstract}
Video is one of the robust sources of information and the consumption of online and offline videos has reached an unprecedented level in the last few years. A fundamental challenge of extracting information from videos is a viewer has to go through the complete video to understand the context, as opposed to an image where the viewer can extract information from a single frame. Apart from context understanding, it almost impossible to create a universal summarized video for everyone, as everyone has their own bias of keyframe, e.g; In a soccer game, a coach person might consider those frames which consist of information on player placement, techniques, etc; however, a person with less knowledge about a soccer game, will focus more on frames which consist of goals and score-board. Therefore, if we were to tackle problem video summarization through a supervised learning path, it will require extensive personalized labeling of data. In this paper, we attempt to solve video summarization through unsupervised learning by employing traditional vision-based algorithmic methodologies for accurate feature extraction from video frames. We have also proposed a deep learning-based feature extraction followed by multiple clustering methods to find an effective way of summarizing a video by interesting key-frame extraction. We have compared the performance of these approaches on the SumMe dataset and showcased that using deep learning-based feature extraction has been proven to perform better in case of dynamic viewpoint videos.
\end{abstract}


%
\IEEEpeerreviewmaketitle

\section{Introduction}
Following the advances of efficient data storage and streaming technologies, videos have become arguably the primary source of information in today's social media-heavy culture and society. Video streaming sites like YouTube are quickly replacing the traditional news and media sharing methods, whom, themselves are forced to adapt the trend of posting videos instead of written articles to convey stories, news and information. This abundance of videos brings forth new challenges of developing an efficient way to extract the subject matter of the videos in question. It would be frustrating, inefficient, unintelligent and downright impossible to watch all movies thoroughly and catalog them according to their categories and subject matter, which is extremely important when searching for a specific video. Currently, this categorization is dependent on the tags, metadata, titles etc., provided by the video uploaders. However, they are highly personalized and unreliable in application. Hence, a better way is required to create a summarized representation of the video that is easily comprehensible in a short amount of time. This is an open research problem in a multitude of fields including information retrieval, networking and computer vision. 

Video Summarization is the process of compacting a video down to only important components in the video. The process is shown in Fig 1. This compact representation is useful in retrieving the desired videos from a large video library. A summarized video must have the following properties:
\begin{itemize}
    \item It must contain the high priority entities and events from the video.
    \item The summary should be free of repetition and redundancy.
\end{itemize}
 Failure to exclude these components might lead to misinterpretation of the video from its summarized version. A summary of video also varies from person to person, so a supervised video summarization can be seen as personalized recommendation approach, but it requires extensive amount of data labeling by a person and yet the trained model isn't a generalized model among mass.

\begin{figure}[ht!]
\centering
  \includegraphics[scale = 0.5]{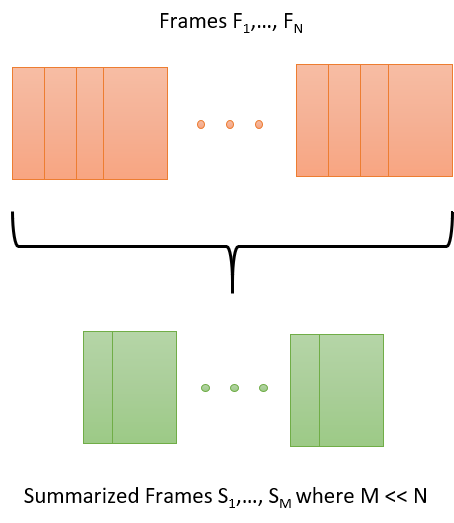}
  \caption{The process of video summarization. N number of frames in the video is summarized to M number of frames where M is far smaller than N.}
\end{figure}

Various approaches have been taken to solve this problem by different researchers. Some of most prominent approaches include keyframe Extraction using Visual Features \cite{gong:2000} \cite{gong2014} and Video Skimming \cite{chorian2013},\cite{Liu2008}. In this paper, we will be exploring a generalized video summarization approach through unsupervised keyframe extraction. We propose a framework to extract vision-based keyframes and use them to create summarized videos with help of clustering and video skimming. We have used the SumMe dataset \cite{GygliECCV14} for our experiments. Our key contributions include suggesting a new unsupervised framework for video summarization. We have first extracted features on basis of traditional computer vision filters and RESNET16 trained on image net, and then used these extracted features for clustering to obtain keyframes. After chosing the keyframes, we just skimmed the video by adding ~1.5 second worth of video around it.

The rest of this paper is organized as follows. The related research is presented in section II, followed by our approach in section III. We present our experimental results is section IV. The paper is concluded with discussions and future goals in section V. Our code is available at  \url{https://github.com/shruti-jadon/Video-Summarization-using-keyframe-Extraction-and-Video-Skimming}

\section{Related Research}
The most difficult challenge of video summarization is determining and separating the important content from the unimportant content. The important content can be classified based on low level features like texture \cite{ahonen2006}, shape \cite{dalal2005} or motion \cite{Bruhn2005}. The frames containing these important information are bundled together to create the summary. This manner of finding key information from static frames is called keyframe extraction. These methods are used dominantly to extract a static summary of the video. Some of the most popular keyframe extraction methods include \cite{ejaz2013}, \cite{lee2012}. These methods use low level features and dissimilarity detection with clustering methods to extract static keyframes from a video. The clustering methods are used to extract the extract features that are worthwhile to be in the summary while uninteresting frames rich with low level features are discarded. Different clustering methods have been used by researchers to find interesting frames \cite{ejaz2013}. Some methods use web-based image priors to extract the keyframes, for example, \cite{kholsa2013}, \cite{Kim2014}.

While extracting static keyframes to compile a summary of the video is effective, the summary itself might not be pleasant to watch and analyze by humans as it will be discontinuous and with abrupt cuts and frame skips. This can be solved by video skimming which appears more continuous and will less abrupt frame changes and cuts. The process is more complex than simple keyframe extraction, however, because a continuous flow of semantic information \cite{jadon2019hands} and relevance is needed to be maintained for videos skimming. Some of the video skimming approaches include \cite{gong:2000}, which utilizes the motion of the camera to extract important information and calculates the inter-frame dissimilarities from the low level features to extract the interesting components from the video. A simple approach to video skimming is to augment the keyframe extraction process by including a continuous set from frames before and after the keyframe up to a certain threshold and include these collection frames in the final summary of the video to create an video skim. 

\begin{figure}[ht!]
\centering
  \includegraphics[scale = 0.22]{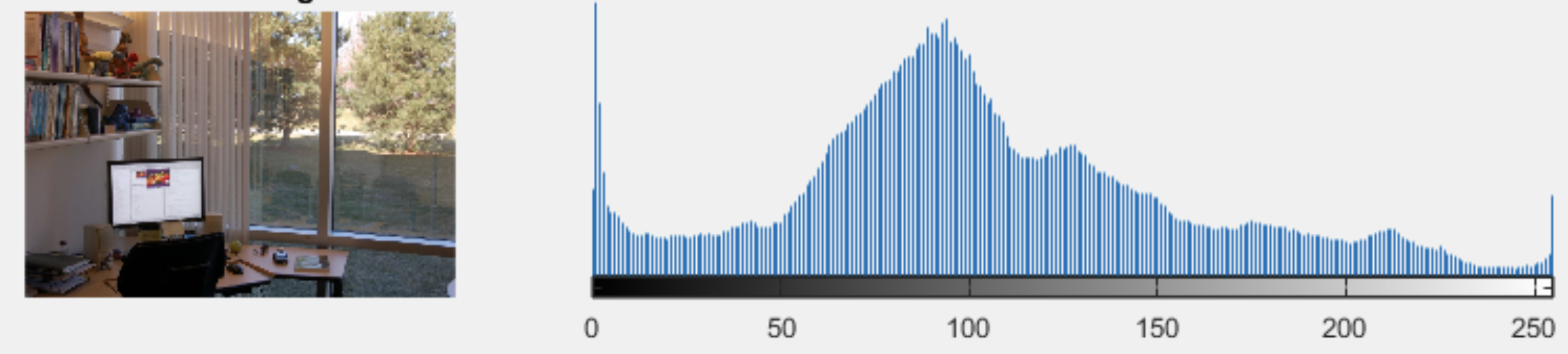}
  \caption{A simple example of Image Histogram.}
\end{figure}

\begin{figure}[ht!]
\centering
  \includegraphics[scale = 0.25]{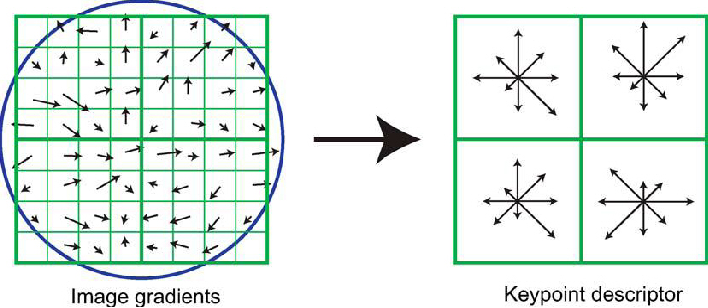}
  \caption{A pictorial representation of SIFT key points extraction from 8X8 Image.}
\end{figure}

\section{Approaches}

In this paper, for static keyframe extraction, we extract low level features using uniform sampling, image histograms, SIFT and image features from Convolutional Neural Network (CNN) trained on ImageNet \cite{ILSVRC15}. We also used two clustering methods: K-means and Gaussian clustering and chosen number of clusters on basis of summarized video size. We have used video skims around the selected keyframes to make the summary fore fluid and comprehensible for humans. We take inspiration from the VSUMM method which is a prominent method in video summarization \cite{deAvila2011} \cite{GygliECCV14}. 

\subsection{keyframe Extraction Techniques}

\subsubsection{Uniform Sampling}
Uniform sampling is one of the most common methods for keyframe extraction \cite{nixon2019feature}. The idea is to select every $k$th frame from the video where the value of $k$ is dictated by the length of the video. A usual choice of length for a summarized video is 5\% to 15\% of the original video, which means every 20th frame in case of 5\% or every 7th frame in case of 15\% length of the summarized video is chosen. For our experiment, we have chosen to use every 7th frame to summarize the video. This is a very simple concept which does not maintain semantic relevance. Uniform sampling is often considered as a generalized baseline for video summarization. 

\begin{figure*}[ht!]
\centering
  \includegraphics[scale = 0.45]{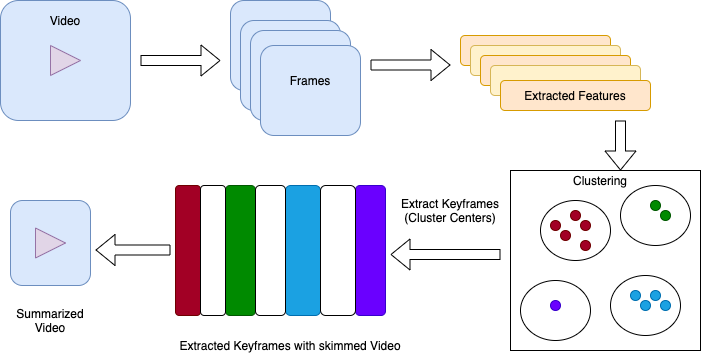}
  \caption{Unsupervised Video Summarization Framework. Here, first video is converted into frames, followed by feature extraction. After feature extraction, they are plotted in an embeddings space, where they are clustered with \% of summarized video as number of clusters. Last, we take center of these clusters as keyframes and apply video skimming to make a continuous summarized video.}
\end{figure*}

\subsubsection{Image histogram}
Image histograms \cite{wang1999image} represent the total distribution of an image. It gives us the number of pixels for a specific brightness values rated from 0 to 256. Image histograms contain important information about images and they can be utilized to extract keyframes. We extract the histogram from all frames. Based on the difference between histograms of two frames, we decide whether the frames have significant dissimilarities among them. We infer that, a significant inter-frame image histogram dissimilarity indicates a rapid change of scene in the video which might contain interesting components. For our experiments, if histograms of two consecutive frames are 50\% or more dissimilar, we extract that frame as a keyframe. 

\subsubsection{Scale Invariant Feature Transform}
Scale Invariant Feature Transform (SIFT) \cite{Morel2010IsT}, has been one of the most prominent local features used in computer vision is applications ranging from object and gesture recognition to video tracking. We use SIFT features for keyframe extraction. SIFT descriptors are invariant to scaling, translation, rotation, small deformations, and partially invariant to illumination, making it a robust descriptor to be used as local features. In SIFT, important locations are first defined using a scale space of smoothed and resized images and applying difference of Gaussian functions on these images to find the maximum and minimum responses. Non maxima suppression is performed and putative matches are discarded to ensure a collection of highly interesting and distinct collection of keypoints. Histogram of oriented gradients is performed by dividing the image into patches to find the dominant orientation of the localized keypoints. These keypoints are extracted as local features. In our experiment, we have extracted HOGs for each frame in video, and then put a threshold which could take 15\% of video. 

\subsubsection{VSUMM} 
This technique has been one of the fundamental techniques in video summarization in the unsupervised setup. The algorithm uses the standard K-means algorithm to cluster features extracted from each frame. Color histograms are proposed to be used in \cite{deAvila2011}. Color histograms are 3-D tensors, where each pixel’s values in the RGB channels determines the bin it goes into. Since each channel value ranges in 0 − 255, usually, 16 bins are taken for each channel resulting in a 16X16X16 tensor. Due to computational reasons, a simplified version of this histogram was computed, where each channel was treated separately, resulting in feature vectors for each frame belonging to R 48 . The nest step suggested for clustering is slightly different. But, the simplified color histograms give comparable performance to the true color histograms. Therefore, we modified VSUMM a bit by using the features extracted from VGG16 at the 2nd fully connected layer \cite{jadon2018introduction}, and clustered it using kmeans.

\subsubsection{ResNet16 on ImageNet}

While reading about approach of VSUMM, we decided to test a different approach. We chose ResNet16 trained on image net, with different range of filters, and chopped of last loss layer, so as to obtain the embeddings of each image (512 dimension). We extracted frames out of the videos, and forward pass them through ResNet16, and after obtaining the embeddings for each frame in video, we clustered them using 2 algorithms: K-means and Gaussian Mixture Models. The number of cluster has been take as 15\% of the video frame numbers. We later chose the frames closest to the center of clusters as the keyframes.

\begin{table*}[t]
  \centering
    \caption{F1 score values of 25 SumMe dataset videos with Human's entry as baseline followed by different approaches of keyframe extraction}
  \begin{tabular}{|c|c|c|c|c|c|c|c|}
  \hline
    Video Name & Human (Avg.) & Uniform Sampling & SIFT & VSUMM(K-means) & VSUMM(Gaussian) & CNN (K-means) & CNN (Gaussian)\\\hline\hline

Base jumping & 0.257 & 0.085364 & 0.234 & 0.083356 & 0.094 & 0.239 & 0.247 \\\hline
Bike Polo & 0.322 & 0.07112 & 0.196 & 0.078369 & 0.065 & 0.204 & 0.212\\\hline
Scuba & 0.217 & 0.0145059 & 0.144 & 0.145599 & 0.172 & 0.195 & 0.184\\\hline
Valparaiso Downhill & 0.217 & 0.19899 & 0.19 & 0.201909 & 0.197 & 0.207 &0.211\\\hline
Bearpark climbing & 0.217 & 0.160377 & 0.146 & 0.156611 & 0.142 & 0.196 & 0.204\\\hline
Bus in Rock Tunnel & 0.217 & 0.030199 & 0.177 & 0.029341 & 0.033 &  0.124 & 0.119\\\hline
Car railcrossing & 0.217 & 0.363804 & 0.36 & 0.386466 & 0.396 & 0.197 & 0.174\\\hline
Cockpit Landing & 0.217 & 0.089413 & 0.035 & 0.906021 & 0.856 & 0.965 & 0.984\\\hline
Cooking & 0.217 & 0.023748 & 0.192 & 0.023172& 0.0257 & 0.205 & 0.197\\\hline
Eiffel Tower & 0.312 & 0.119034 & 0.004 & 0.123115 &0.135 & 0.157 & 0.146\\\hline
Excavators river crossing & 0.303 & 0.328008 & 0.32 & 0.326871 & 0.345 & 0.342 & 0.357\\\hline
Jumps & 0.483 & 0.176244 & 0.16 & 0.174919 & 0.185 & 0.182 & 0.176\\\hline
Kids playing in leaves & 0.289 & 0.426775 & 0.366 &0.424418 & 0.482 & 0.372 & 0.384\\\hline
Playing on water slide & 0.195 & 0.168675 & 0.232 & 0.174321 & 0.185 & 0.278 & 0.297\\\hline
Saving dolphines &0.188 &  0.212642 & 0.121 & 0.229369 & 0.257 & 0.247 & 0.217\\\hline
St Maarten Landing & 0.496 & 0.0404343 & 0.12 & 0.039482 & 0.0254 & 0.059 & 0.068\\\hline
Statue of Liberty & 0.184 & 0.068651 & 0.208 & 0.070949 & 0.072 & 0.095 &0.097\\\hline
Uncut Evening Flight & 0.35 & 0.253156 & 0.256 & 0.251676 & 0.274 & 0.278 &0.295\\\hline
paluma jump & 0.509 & 0.048565 & 0.092 & 0.047268 & 0.048 &0.049 &0.049\\\hline
playing ball & 0.271 & 0.239955 & 0.222 & 0.258244 & 0.237 & 0.256 & 0.258\\\hline
Notre Dame & 0.231 & 0.229265 & 0.23 &0.223917 & 0.021 & 0.0230 & 0.0227\\\hline
Air Force One & 0.332 & 0.066812 & 0.07 & 0.065103 & 0.061 &0.065 & 0.048\\\hline
Fire Domino & 0.394 & 0.002603 & 0.247 & 0.003367 & 0.0020 & 0.0042 & 0.0035\\\hline
car over camera & 0.346 & 0.035693 & 0.04 & 0.038304 & 0.035 & 0.0458 & 0.0475\\\hline
Paintball & 0.399 & 0.224322 & 0.23 & 0.233006 & 0.245 & 0.297 & 0.304\\\hline
mean & 0.311 & 0.0152 & 0.171 & 0.155 &0.1869 &0.1765 &0.212 \\\hline
  \end{tabular}
  \label{tab:1}
\end{table*}

\subsection{Clustering}
\subsubsection{K-means clustering}
K-means clustering is a very popular clustering method. Given a set of image frames extracted by one of the methods mentioned in section III-A, the goal is to partition these frames into different clusters, so that the within-cluster sum of squared difference is minimum. This is equivalent to minimizing the pairwise squared deviation of points in the same cluster. With this clustering we find the interesting frames to be included in the summarization and discard the ones that are rich in local features but contains less informative or interesting content. \\
For our paper, we have used Kmeans for clustering the features obtained from RESNET16 ImageNet trained method. We obtained 512 dimension vector for each frame in video, and clustered them. We have set the number of cluster to be 15\% of the video. After clustering, we chose the key points which was closest to the center of that specific cluster. 

\subsubsection{Gaussian Clustering (Mixture Model)}
Gaussian mixture models (GMM) are often used for data clustering. Usually, fitted GMMs cluster by assigning query data points to the multivariate normal components that maximize the component posterior probability given the data. That is, given a fitted GMM, a cluster assigns query data to the component yielding the highest posterior probability. This method of assigning a data point to exactly one cluster is called hard clustering. \\
However, GMM clustering is more flexible because you can view it as a fuzzy or soft clustering method. Soft clustering methods assign a score to a data point for each cluster. The value of the score indicates the association strength of the data point to the cluster. As opposed to hard clustering methods, soft clustering methods are flexible in that they can assign a data point to more than one cluster. \\
In this paper, we used clustering on the embeddings obtained using RESNET16 trained network. we set the number of clusters to be 15\% of the video, then chose the points which were closest to the center of the cluster.

\subsection{Video Summarization}
Our approach for video summarization is influenced by the VSUMM method \cite{deAvila2011}. Firstly, keyframes containing important information is extracted using one of the methods mentioned in section III-A. To reduce the computation time for video segmentation, a fraction of the frames were used. Considering the sequence of frames are strongly correlated, the difference from one frame to the next is expected to be very low when sampled at high frequencies, such as, 30 frames per second. Instead using a low frequency rate of 5 frames per second had insignificant effect on the results but it increased the computation speed by a significant margin. We used 5 frames per second as a sampling rate for our experiments and discarded the redundant frames. 

After extracting all the keyframes, we perform a clustering on the frames to categorized them into interesting and uninteresting frames using one of the methods mentioned in section III-B. The cluster with the interesting frames were used to generate the summary of the video. The summary of the video was chosen to have the length of approximately 15\% of the original video. But this summary was discontinuous and thus different from the way a human observer would evaluate the summary leading to poor scores as our evaluation method coincides with how a human being scores the summary. This problem was overcome by using a 1.8 second skims from the extracted interesting frame. This makes the summary continuous and easy to comprehend. The low frequency sampling of frames helps keep the size if the video in check. 

\section{Experiments and Results}

\subsection{Dataset}

For our experimentation, we use the SumMe dataset \cite{Gygli2014} which was created to be used as a benchmark for video summarization. The dataset contains 25 videos with the length ranging from one to six minutes. Each of these videos are annotated by at least 15 humans with a total of 390 human summaries. The annotations were collected by crowd sourcing. The length of all the human generated summaries are restricted to be within 15\% of the original video. Frames from two example videos, a) Air Force One and b) Play Ball is presented in Fig 3. 

\subsection{Evaluation Method}

The SumMe dataset provides individual scores to each annotated frames. We evaluate our method by measuring the F-score from the set of frames that have been selected by our method. We compare the F-score to the human generated summaries to validate the effectiveness of our method. 
F-score is a measure that combines precision and recall is the harmonic mean of precision and recall, the traditional F-measure or balanced F-score:

${\displaystyle F=2\cdot {\frac {\mathrm {precision} \cdot \mathrm {recall} }{\mathrm {precision} +\mathrm {recall} }}} {\displaystyle F=2\cdot {\frac {\mathrm {precision} \cdot \mathrm {recall} }{\mathrm {precision} +\mathrm {recall} }}}$

This measure is approximately the average of the two when they are close, and is more generally the harmonic mean, which, for the case of two numbers, coincides with the square of the geometric mean divided by the arithmetic mean. There are several reasons that the F-score can be criticized in particular circumstances due to its bias as an evaluation metric. This is also known as the $ F_{1}$ measure, because recall and precision are evenly weighted.

\begin{figure*}[ht]
\centering
  \includegraphics[width=\textwidth]{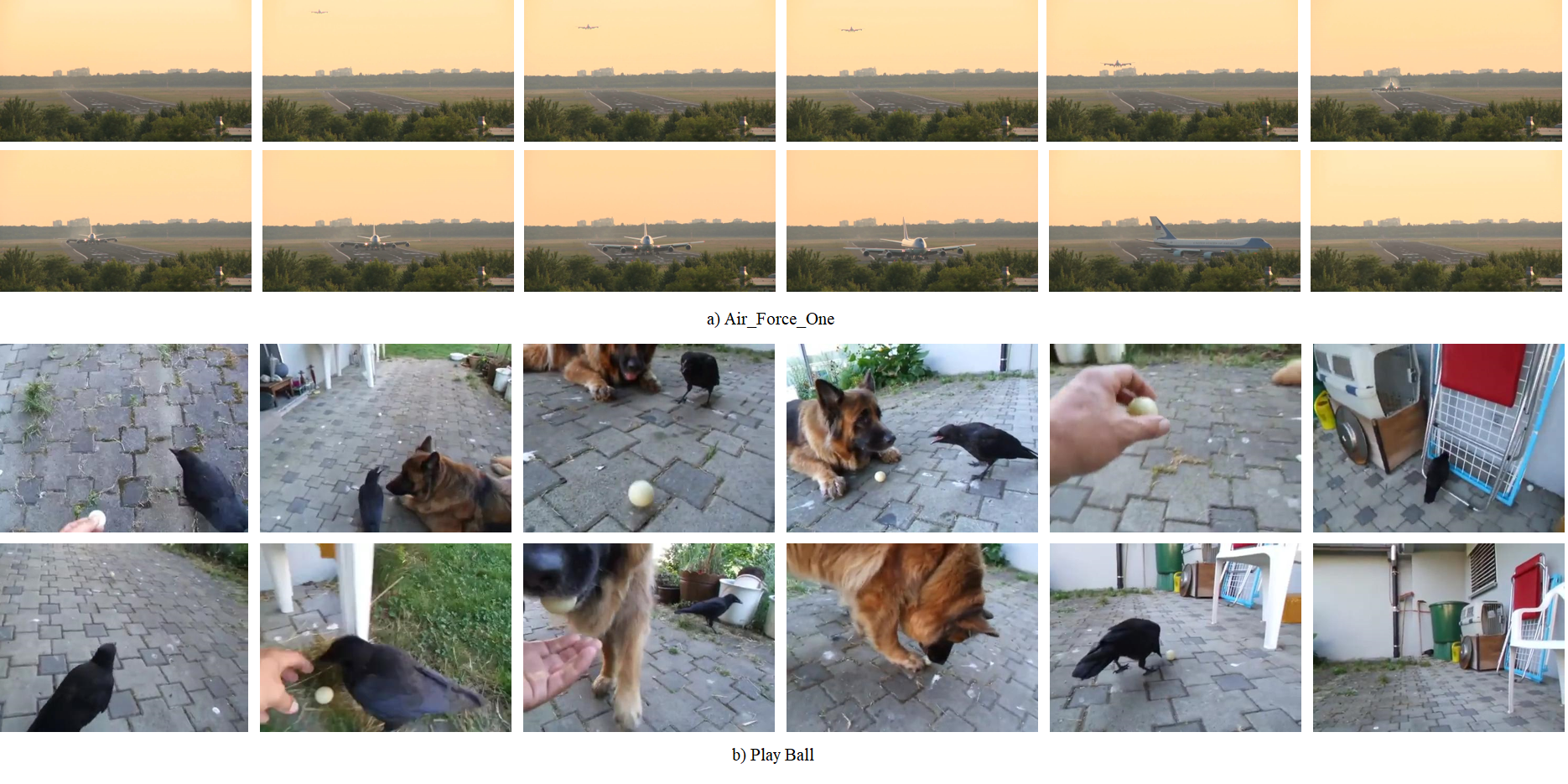}
  \caption{Two example videos from the SumMe dataset. a) Air Force One has a fixed camera and mostly static background. b) Play Ball - has moving camera with dynamic background.}
\end{figure*}

\subsection{Results}
We ran mentioned methods on the SumMe Dataset, and compared the F-scores obtained by them (as shown in Table 1) .Our main goal is to be as much close to human, which we were able to obtain using SIFT, VSUMM, and CNN. We also took mean of scores for all videos, and can see that CNN(Gaussian) was performing good followed by VSUMM. We observed that, the videos which had dynamic view point was performing good with VSUMM and CNN, whereas the videos with stable view point was performing very poor even with compared to Uniform Sampling. This is where we can find difference in a human's method of summarizing vs an algorithm method.We can also see that SIFT's and CNN's have positive correlation in terms of F-scores this is due to the features obtained. Though, SIFT is not able to outperform CNN. 

\section{Conclusion}
Video Summarization is one of the hardest task because it depends on person's perception. So, we can never have a good baseline to understand whether our algorithm is working or not. Sometimes, Humans just want 1-2 second of video as summary, whereas machine looks for slightest difference in image intensity and might give us 10 seconds of video. 

From what the baseline has been given in SumMe Dataset, we chose the average human baseline as true, as we would like to consider all perspectives. After testing with all different forms of videos, we can conclude that Gaussian Clustering along with Convolutional Networks can give better performance than other methods with moving point camera videos. In fact, the SIFT algorithm seems to perform well on videos with high motion, the reason behind it is that we used deep layered features, thus they consists of important points inside image, followed by Gaussian Clustering, which is specifically made for mixture based components. We have also observed that, even Uniform Sampling is giving better result for videos which have stable camera view point and very less motion.\\
We can conclude that one single algorithm can't be solution of video summarization, it is dependent of the type of video, the motion inside video.

\ifCLASSOPTIONcaptionsoff
  \newpage
\fi

\bibliographystyle{IEEEtran}
\bibliography{main}

\end{document}